\documentclass[journal,12pt,onecolumn]{IEEEtran}

\usepackage{setspace}
\doublespace
\usepackage{graphics}
\usepackage{rotating}
\usepackage{amsfonts}
\usepackage{amsmath}
\usepackage{subfigure}
\usepackage{comment}
\usepackage{multirow}
\allowdisplaybreaks[1]
\begin{document}
\title{\large\bf
Computational Complexity of Decoding
Orthogonal Space-Time Block Codes}
\author{
\normalsize
Ender Ayanoglu$^*$, Erik G. Larsson$^\dag$, and Eleftherios Karipidis$^\dag$
\thanks{$^*$E. Ayanoglu is with the Center for Pervasive Communications and
  Computing, 
  University of California Irvine, Irvine, CA 92697-2625.}
\thanks{$^\dag$E. G. Larsson and E. Karipidis are with the Department of
  Electrical Engineering, Link\"oping University, SE-581 83
  Link\"oping, Sweden.} 
}
\maketitle
\vspace{-0.1in}
\begin{abstract}
The computational complexity of optimum decoding for an orthogonal
space-time block code ${\cal G}_N$ satisfying 
${\cal G}_N^H{\cal G}_N=c(\sum_{k=1}^K|s_k|^2)I_N$ where $c$ is a
positive integer is quantified. Four equivalent techniques of
optimum decoding which have the same computational complexity are
specified. Modifications to the basic formulation in special cases
are calculated and illustrated by means of examples. 
This paper 
corrects and extends \cite{aa},\cite{aa2}, and unifies them with the
results from the literature. In addition, a number of results from 
the literature are extended to the case $c>1$.
\end{abstract}
\section{Introduction}
In \cite{tjc2}, an optimum Maximum Likelihood metric is 
introduced for Orthogonal Space-Time Block Codes (OSTBCs).
A general description of this metric and specific
forms for a number of space-time codes can be found in \cite{vy}. 
This metric is complicated and, in a straightforward implementation, its
computational complexity would depend on the size of the signal
constellation. By a close inspection, it can be observed that it
can actually be simplified and made independent of the constellation size.
Alternatively, the Maximum Likelihood
formulation can be made differently and the simplified metric can be
obtained via different formulations \cite{ls},\cite{glmz}. In
\cite{aa},\cite{aa2}, yet another formulation is provided. 
In this paper, we will unify all of the approaches cited above and 
calculate the computational complexity of the optimum decoding of an
OSTBC. We will begin our discussion within the framework of 
\cite{aa},\cite{aa2}. 

Consider the decoding of an OSTBC
with
$N$ transmit and $M$ receive antennas, and an interval of $T$ symbols
during which the channel is constant. The received
signal is given by 
\begin{equation}
Y={\cal G}_N H+V
\label{eq-1}
\end{equation}
where $Y=[y_t^j]_{T\times M}$ is the received signal matrix of size
$T\times M$ and whose entry $y_t^j$ is the signal received at antenna
$j$ at time $t$, $t=1,2,\ldots,T$, $j=1,2\ldots,M$; $V=[v_t^j]_{T\times
  M}$ is the noise matrix, and ${\cal G}_N=[g_t^i]_{T\times N}$ is the
transmitted signal matrix whose entry $g_t^i$ is the signal
transmitted at antenna $i$ at time $t$, $i=1,2,\ldots,N$. The matrix
$H=[h_{i,j}]_{N\times M}$ is the channel coefficient matrix of size
$N\times M$ whose entry $h_{i,j}$ is the channel coefficient from
transmit antenna $i$ to receive antenna $j$. The entries of the
matrices $H$ and $V$ are independent, zero-mean, and circularly
symmetric complex Gaussian random variables. ${\cal G}_N$ is an OSTBC
with complex symbols $s_k$, $k=1,2,\ldots,K$ and therefore
${\cal G}_N^H{\cal G}_N=c(\sum_{k=1}^K|s_k|^2)I_N$ where $c$ is a positive
integer and $I_N$ is the identity matrix of size $N$. 
\section{A Real-Valued Representation}
Arrange the matrices $Y$, $H$, and $V$, each in one column
vector by stacking their columns on top of one another
\begin{eqnarray}
y & = & {\rm vec}(Y)=(y_1^1,\ldots,y_T^M)^T,\\
h & = & {\rm vec}(H) =(h_{1,1},\ldots,h_{N,M})^T,\\
v & = & {\rm vec}(V)=(v_1^1,\ldots,v_T^M)^T.
\end{eqnarray}
Then one can write
\begin{equation}
y = \check {\cal G}_N h + v
\label{eq-2}
\end{equation}
where $\check{\cal G}_N = I_M \otimes {\cal G}_N$, with 
$\otimes$ denoting the Kronecker
matrix multiplication.
In \cite{aa},\cite{aa2}, 
a real-valued representation of (\ref{eq-1}) is obtained by
decomposing the $MT$-dimensional
complex problem defined by (\ref{eq-2}) to a $2MT$-dimensional
real-valued problem by applying the real-valued lattice representation
defined in \cite{aa3} to obtain 
\begin{equation}
\check y = \check H x + \check v
\label{eq-3}
\end{equation}
where
\begin{eqnarray}
\check y & = & ({\rm Re}(y_1^1),{\rm Im}(y_1^1),\ldots,
{\rm Re}(y_T^M),{\rm Im}(y_T^M))^T,\\
x & = & ({\rm Re}(s_1),{\rm Im}(s_1),\ldots,
{\rm Re}(s_K),{\rm Im}(s_K))^T,\\
\check v & = & ({\rm Re}(v_1^1),{\rm Im}(v_1^1),\ldots,
{\rm Re}(v_T^M),{\rm Im}(v_T^M))^T .
\end{eqnarray}
The real-valued fading coefficients of $\check H$ are defined using
the complex fading coefficients $h_{i,j}$ from transmit antenna $i$ to
receive antenna $j$ as $h_{2i-1+2(j-1)N}={\rm Re}(h_{i,j})$ and 
$h_{2i+2(j-1)N}={\rm Im}(h_{i,j})$ for $i=1,2,\ldots,N$ and
    $j=1,2,\ldots,M$. Since ${\cal G}_N$ is an orthogonal matrix and
    due to the real-valued representation of the system using
    (\ref{eq-3}), it can be observed that the columns $\check h_i$ of
$\check H$ are orthogonal to each other and their inner products with
    themselves are a constant \cite{aa},\cite{aa2}
\begin{equation}
\check H^T \check H = \sigma I_{2K}.
\label{eq-8}
\end{equation}
By multiplying (\ref{eq-3}) by $\check H^T$ on the left, we have 
\begin{equation}
\bar{\bar y} = \sigma x + \bar{\bar v}
\label{eq-11}
\end{equation}
where $\bar{\bar y}=\check H^T\check y$, and
$\bar{\bar v}=\check H^T\check y$ is a zero-mean random vector. Due to
(\ref{eq-8}), $\bar{\bar v}$
has independent and identically distributed Gaussian members. The Maximum
Likelihood solution is found by minimizing

\begin{equation}
\|\bar{\bar y}-\sigma x\|_2^2
\label{eq-12}
\end{equation}
or equivalently
\begin{equation}
\|\sigma^{-1}\bar{\bar y}- x\|_2^2
\end{equation}
over all combinations of $x\in \Omega^{2K}$. 
As a result, the joint detection problem of an OSTBC decouples into
$K$ symbol detection problems
\begin{equation}
\|\sigma^{-1}(\bar{\bar y}_{2k-1},\bar{\bar y}_{2k})-(x_{2k-1},x_{2k})\|_2^2
\end{equation}
one per symbol $(x_{2k-1},x_{2k})\in \Omega^2$, where
$k=1,2,\ldots,K$. Further, we assume that the signal constellation is 
separable as $\Omega^2$ where 
$\Omega=\{\pm 1,\pm 3\ldots,\pm (2L-1)\}$, and $L$
is an integer, the Maximum Likelihood decoding problem can be further
simplified to
\begin{equation}
\min_{x_k\in \Omega} |\hat x_k - x_k |^2
\label{eq-13}
\end{equation}
where we denoted
\begin{equation}
\hat x_k = \sigma^{-1}\bar{\bar y}_k, \hspace{8mm}k=1,2,\ldots,2K,
\label{eq-111}
\end{equation}
which is a standard operation in conventional Quadrature Amplitude
Modulation (QAM). In the sequel, we
will compute the decoding complexity up to this quantization
operation.

The decoding operation consists of the multiplication 
\begin{equation}
\bar{\bar y} = \check H^T\check y, 
\label{eq-10}
\end{equation}
the calculation of 
\begin{equation}
\sigma = \check h_1^T\check h_1,
\label{eq-6}
\end{equation}
the inversion of $\sigma$, and the multiplications in
(\ref{eq-111}). 

In what follows, we will show that when 
${\cal G}_N^H{\cal G}_N=c(\sum_{k=1}^K |s_k|^2)I_N$ where $c$ is a
positive integer, then 
$\sigma=c\|H\|^2$. The development will lead to the four equivalent
optimal decoding techniques discussed in the next section.

Let $\bar s_k ={\rm Re}[s_k]$ and $\tilde s_k = {\rm Im}[s_k]$.
Form two vectors, $\bar s$ and $\tilde s$,
consisting of $\bar s_k$ and $\tilde s_k$, respectively, and
form a vector $s'$ that is the concatenation of $\bar s$ and $\tilde s$
\begin{equation}
\bar s = (\bar s_1 , \bar s_2 , \ldots , \bar s_K)^T,\hspace{4mm}
\tilde s = (\tilde s_1 , \tilde s_2 , \ldots , \tilde
s_K)^T,\hspace{4mm}
s'=(\bar s^T , \tilde s^T)^T.
\label{eq-42}
\end{equation}
By rearranging the right hand side of (\ref{eq-2}), we can write
\begin{equation}
y = F s' + v = F_a \bar s + F_b \tilde s + v
\label{eq-43}
\end{equation}
where $F=[F_a \hspace{1mm}F_b]$ is an $MT\times 2K$ complex matrix and 
$F_a$ and $F_b$ are $MT\times K$ complex matrices whose entries
consist of (linear 
combinations of) channel coefficients $h_{i,j}$. In \cite{ls}, it was
shown that when ${\cal G}_N^H{\cal G}_N=(\sum_{k=1}^K |s_k|^2)I_N$, then
${\rm Re} [ F^HF]=\| H\|^2 I_N$. It is straightforward to extend this result so
that when ${\cal G}_N^H{\cal G}_N=c (\sum_{k=1}^K |s_k|^2)I_N$, then 
\begin{equation}
{\rm Re} [ F^HF]=c \| H\|^2 I
\label{eq-44}
\end{equation}
where $c$ is a positive integer. Let 
\begin{equation}
\bar y = {\rm Re}[y], \hspace{4mm}\tilde y = {\rm Im}[y],\hspace{4mm}
\bar v = {\rm Re}[v], \hspace{4mm}\tilde v = {\rm Im}[v],
\label{eq-45}
\end{equation}
and
\begin{equation}
\bar F_a={\rm Re}[F_a],\hspace{4mm}\tilde F_a={\rm Im}[F_a],\hspace{4mm}
\bar F_b={\rm Re}[F_b],\hspace{4mm}\tilde F_b={\rm Im}[F_b].
\label{eq-46}
\end{equation}
Now define
\begin{equation}
y'=\left[\begin{array}{c}\bar y\\\tilde y \end{array}\right]
\hspace{4mm}
F'=\left[\begin{array}{cc}\bar F_a & \bar F_b\\\tilde F_a &\tilde F_b
\end{array}\right]\hspace{4mm}
v'=\left[\begin{array}{c}\bar v\\\tilde v\end{array}\right]
\hspace{4mm}
\label{eq-47}
\end{equation}
so that we can write
\begin{equation}
y' = F' s' + v'
\label{eq-48}
\end{equation}
which is actually the same expression as (\ref{eq-3}) except
the vectors and matrices have their rows and columns permuted. 

It can
be shown that (\ref{eq-44}) implies
\begin{equation}
F'\hspace{.5mm}^TF'=c\|H\|^2 I.
\label{eq-49}
\end{equation}

Let $P_y$ and $P_s$ be $2MT\times 2MT$ and $2K\times 2K$,
respectively, permutation matrices such that 
\begin{equation}
\check y =P_yy',\hspace{4mm} x = P_ss' . 
\label{eq-50}
\end{equation}
It follows that $P_y^TP_y^{\ }=P_y^{\ }P_y^T=I$ and 
$P_s^TP_s^{\ }=P_s^{\ }P_s^T=I$.
We now have
\begin{equation}
\check y = 
P_y(F's'+v') = P_yF'P_s^T x +P_yv' = \check H x + \check v .
\label{eq-53}
\end{equation}
Therefore,
\begin{equation}
\check H = P_yF'P_s^T
\label{eq-54}
\end{equation}
which implies
\begin{equation}
\check H^T\check H = P_s^{\ }F'\hspace{0.5mm}^TP_y^TP_y^{\ }F'P_s^T
=c \| H \|^2 I.
\label{eq-55}
\end{equation}
As a result, $\sigma=c\| H\|^2$.
%
\section{Four Equivalent Optimum Decoding Techniques for OSTBCs}
For an OSTBC ${\cal G}_N$ satisfying 
${\cal G}_N^H{\cal G}_N=c(\sum_{k=1}^K|s_k\|^2)I_N$ where $c$ is a
positive integer, the Maximum Likelihood solution is formulated in
four equivalent ways with equal squared distance values
\begin{equation}
\|Y-{\cal G}_NH\|^2=\|y-Fs'\|^2=\|y'-F's'\|^2=\|\check y-\check Hx\|^2.
\label{eq-57}
\end{equation}
There are four solutions, all equal. The first solution is obtained by
expanding $\|Y-{\cal G}_NH\|^2$ and is given by eq. (7.4.2) of
\cite{ls} when $c=1$\footnote{The notation in \cite{vy} and \cite{ls} is the
transposed form of the one adopted in this paper.}. When $c>1$, it
should be altered as 
\begin{equation}
\hat s_k = \frac{1}{c \| H\|^2}[{\rm Re}\{{\rm Tr}(H^HA_k^HY)\}-\hat\imath\cdot
  {\rm Im}\{{\rm Tr}(H^HB_k^HY)\}] \hspace{1cm}k=1,2,\ldots,K
\label{eq-58}
\end{equation}
where $A_k$ and $B_k$ are the matrices in
the linear 
representation of ${\cal G}_N$ in terms of $\bar s_k$
and $\tilde s_k$ for $k=1,2,\ldots,K$ as 
\begin{equation}
{\cal G}_N=\sum_{k=1}^K{\bar s_k}A_k+\hat\imath {\tilde s_k}B_k 
= \sum_{k=1}^K s_k{\check A}_k+s_k^*{\check B}_k,
\label{eq-21}
\end{equation}
$\hat\imath = \sqrt{-1}$, $A_k=\check A_k+\check B_k$, and $B_k=\check
A_k-\check B_k$ \cite{ls}. 
Once $\{\hat s_k\}_{k=1}^K$ are calculated, the decoding problem can
be solved by
\begin{equation}
\min_{\bar s_k\in \Omega}|\bar s_k-{\rm Re}[\hat s_k]|^2,
\hspace{8mm}
\min_{\tilde s_k\in \Omega}|\tilde s_k-{\rm Im}[\hat s_k]|^2
\label{eq-22}
\end{equation}
once for each $k=1,2,\ldots,K$. Similarly to (\ref{eq-13}), this is a
standard quantization problem in QAM.

The second solution is obtained by expanding the second expression in
(\ref{eq-57}) and is given by 
\begin{equation}
\hat s'=\frac{{\rm Re}[F^Hy]}{c\| H\|^2} .
\label{eq-59}
\end{equation}
This is given in [4. eq. (7.4.20)] for $c=1$. The third solution 
corresponds to the minimization of the third expression in (\ref{eq-57}) 
and is given by
\begin{equation}
\hat s'=\frac{F'\hspace{.5mm}^T y'}{c \|H\|^2} .
\label{eq-60}
\end{equation}
The fourth solution is the one introduced in \cite{aa}. It is obtained
by minimizing the fourth expression in (\ref{eq-57}) and is given by
\begin{equation}
\hat x =\frac{\check H^T \check y}{\sigma}
=\frac{\check H^T \check y}{c\| H\|^2} .
\label{eq-61}
\end{equation}
Considering that
\begin{equation}
F_a=[{\rm vec}(A_1H)\hspace{1mm}\cdots \hspace{1mm}{\rm vec}(A_KH)]\hspace{4mm}
F_b=[\hat\imath\cdot {\rm vec}(B_1H)\hspace{1mm}\cdots
  \hspace{1mm}\hat\imath\cdot {\rm vec}(B_KH)] 
\end{equation}
[4, eq. (7.1.7)], it can be verified that (\ref{eq-58}) and 
(\ref{eq-59}) are equal. The equality of (\ref{eq-59}) and
(\ref{eq-60}) follows from (\ref{eq-45})-(\ref{eq-47}). The equality of
(\ref{eq-60}) and (\ref{eq-61}) follows from (\ref{eq-50}) and (\ref{eq-54}).
Therefore, equations (\ref{eq-58}), (\ref{eq-59})-(\ref{eq-61}) yield the same
result, and when 
properly implemented, will have identical computational
complexity. 

Although these four techniques are equivalent, a straightforward
implementation of (\ref{eq-58}) or (\ref{eq-59}) 
can actually result in larger complexity than (\ref{eq-60}) or
(\ref{eq-61}). The proper implementation requires that in
(\ref{eq-58}) 
or (\ref{eq-59}), the terms not needed due to elimination by the Tr[ ],
Re[ ], and Im[ ] operators are not calculated. 

Let's now compare these techniques with the minimization of the metric
introduced in \cite{tjc2}. For a complex OSTBC, let \cite{tjc2},\cite{vy}
\begin{equation}
r_k = \sum_{t\in\eta(k)}\sum_{j=1}^M{\rm sgn}_t(k)
\breve h_{\epsilon(k),j}\breve y_t^j(k)
\label{eq-101}
\end{equation}
where $\eta(k)$ is the set of rows of ${\cal G}_N$ in which $s_k$
appears, $\epsilon_t(k)$ expresses the column position of $s_k$ in the
$t$th row, ${\rm sgn}_t(k)$ denotes the sign of $s_k$ in the $t$th
row,
\begin{equation}
\breve h_{\epsilon_t(k),j}=\left\{\begin{array}{ll}
h_{\epsilon_t(k),j}^*
& {\rm if\ }s_k{\rm\ is\ in\ the\ }t{\rm th\ row\ of\ }{\cal G}_N,\\
h_{\epsilon_t(k),j} ^{}
& {\rm if\ }s_k^*{\rm\ is\ in\ the\ }t{\rm th\ row\ of\ }{\cal G}_N,
\end{array}\right.
\end{equation}
and
\begin{equation}
\breve y_t^j(k)=\left\{\begin{array}{ll}
y_t^j & {\rm if\ }s_k{\rm\ is\ in\ the\ }t{\rm th\ row\ of\ }{\cal G}_N,\\
(y_t^j)^* & {\rm if\ }s_k^*{\rm\ is\ in\ the\ }t{\rm th\ row\ of\ }{\cal G}_N
\end{array}\right.
\label{eq-102}
\end{equation}
for $k=1,2,\ldots,K$. A close inspection shows that $r_k$ in
(\ref{eq-101})-(\ref{eq-102}) is equal to the numerator of (\ref{eq-58}).

The metric to be minimized for $s_k$ is
given as \cite{tjc2},\cite{vy}
\begin{equation}
|s_k- r_k|^2+\left(c\sum_{i=1}^N\sum_{j=1}^M|h_{i,j}|^2-1
\right)|s_k|^2.
\label{eq-103}
\end{equation}
Implemented as it appears in (\ref{eq-103}), this metric has larger
complexity than the metrics for four equivalent techniques described
above. Furthermore, its complexity depends on the constellation size
$L$ due to the presence of the factor $|s_k|^2$. It can be simplified,
however. 

For minimization purposes, we can write (\ref{eq-103}) as 
\begin{align}
|s_k|^2  -  2{\rm Re}[s_k^* r_k] +|r_k|^2+c\|H\|^2 & |s_k|^2-  |s_k|^2\notag\\
& =  c\|H\|^2\left(|s_k|^2-\frac{2{\rm Re}[s_k^* r_k]}{c\|H\|^2}
+ \frac{| r_k|^2}{c^2\|H\|^4}\right)+{\rm const.}\label{eq-104} \\
& =  c\|H\|^2\left|s_k-\frac{ r_k}{c\|H\|^2}\right|^2+{\rm const.}\notag
\end{align}
where the first equality follows from the fact that the third term
inside the paranthesis in (\ref{eq-104}) is independent of
$s_k$. Because of our observation that $ r_k$ is the same as
the numerator of (\ref{eq-58}), we have
\begin{equation}
\hat s_k = \frac{r_k}{c\|H\|^2}\hspace{4mm}k=1,2,\ldots,K
\end{equation}
and then this method becomes equivalent to our four equivalent techniques.
\vspace{-3mm}
\section{Optimum Decoding Complexity of OSTBCs}
Since the four decoding techniques
(\ref{eq-58}), (\ref{eq-59})-(\ref{eq-61}) are equivalent, we will
calculate their computational complexity by using one of them. 
This can be done most simply by using (\ref{eq-60}) or (\ref{eq-61}).
We will use (\ref{eq-61}) for this purpose. 


First, assume $c=1$. Note $\check H$ is a $2MT\times 2K$ matrix. 
The multiplication $\check H^T\check y$ takes
$2MT\cdot 2K$ and calculation of $\sigma=\|H\|^2$ takes $2MN$ real
multiplications, its 
inverse takes a real division, and $\sigma^{-1}\bar{\bar y}$ takes $2K$ real
multiplications. Similarly, the multiplication $\check H^T\check y$
takes $2K \cdot (2MT-1)$, and calculation of $\sigma$ takes $2MN-1$
real additions. Letting $R_D$, $R_M$ and $R_A$ be the number of real 
divisions, the number of real 
multiplications, and the number of real additions, the complexity of
decoding the transmitted complex signal $(s_1,s_2,\ldots,s_K)$ with
the technique described in (\ref{eq-10}),(\ref{eq-6}),(\ref{eq-111}) is
\begin{equation}
{\cal C}=1 R_D, (4KMT+2MN+2K)R_M,(4KMT+2MN-2K-1)R_A.
\label{eq-17-}
\end{equation}
Note that the complexity does not depend on the constellation size $L$. 
If we take the complexity of a real division as equivalent to 4 real
multiplications as in \cite{aa},\cite{aa2}, then the complexity is
\begin{equation}
{\cal C}=(4KMT+2MN+2K+4)R_M,(4KMT+2MN-2K-1)R_A
\label{eq-17}
\end{equation}
which is smaller than the complexity specified in \cite{aa},\cite{aa2}
and does not depend on $L$.
In the rest of this
paper, we will use this assumption. The conversion from this form to
that in (\ref{eq-17-}) can be made simply by adding a real division and
reducing the number of real multiplications by 4.

When $c>1$, the number of real multiplications to calculate $\sigma$
increases by 1, however, in the examples it will be seen that 
the complexity of the calculation of $\check
H^T\check y$ is reduced by a factor of $c$. 

In what follows, 
we will calculate the exact complexity values for four examples.
See \cite{tjc2},\cite{vy} for explicit metrics of the form
(\ref{eq-101})-(\ref{eq-103}) for these examples.

{\em Example 1:\/} Consider the Alamouti OSTBC with $N=K=T=2$ and $M=1$
where 
\begin{equation}
{\cal G}_2 = \left [\begin{array}{cc}s_1&s_2\\ -s_2^*&s_1^*\\
\end{array}\right ] .
\label{eq-23}
\end{equation}
The matrix $\check H$ can be calculated as
\begin{equation}
\check H = \left[\begin{array}{cccc}
h_1&-h_2&h_3&-h_4\\
h_2&h_1&h_4&h_3\\
h_3&h_4&-h_1&-h_2\\
h_4&-h_3&-h_2&h_1\\
\end{array}\right] .
\label{eq-26}
\end{equation}
Note that the matrix $\check H$ is orthogonal and all of its columns
have the same squared norm. One needs 16 real multiplications 
to calculate $\bar{\bar y}=\check H^T\check y$, 4 real multiplications
to calculate 
$\sigma=\check h_1^T\check h_1$, 4 real multiplications to calculate
$\sigma^{-1}$, and 4 real multiplications to calculate $\sigma^{-1}\bar
y$. There are $3\cdot 4=12$ real
additions to calculate $\check H^T\check y$ and 3 real additions to
calculate $\sigma$. As a result, with this approach, decoding takes a total
of 28 real multiplications and 15 real additions. 

The complexity figures in (\ref{eq-17}) 
are 28 real multiplications and 15 real additions, which hold exactly.

{\em Example 2:\/} Consider the OSTBC with $M=2$, $N=3$, $T=8$, and
$K=4$ given by \cite{tjc} 
\begin{equation}
{\cal G}_3=\left[
\begin{array}{cccccccc}
s_1&-s_2&-s_3&-s_4&s_1^*&-s_2^*&-s_3^*&-s_4^*\\
s_2&s_1&s_4&-s_3&s_2^*&s_1^*&s_4^*&-s_3^*\\
s_3&-s_4&s_1&s_2&s_3^*&-s_4^*&s_1^*&s_2^*\\
\end{array}
\right]^T .
\label{eq-28}
\end{equation}
For this ${\cal G}_N$, one has 
$
{\cal G}_3^H{\cal G}_3=2\left ( \sum_{k=1}^K|s_k|^2\right ) I_3.
$
In \cite{aa2}, it has been shown that the $32\times 8$ real-valued
channel matrix $\check H$ is
\begin{equation}
\check H = \left [
\begin{array}{cccccccc}
h_1 & -h_2 & h_3 & -h_4 & h_5 & -h_6 & 0 & 0 \\
h_2 & h_1 & h_4 & h_3 & h_6 & h_5 & 0 & 0 \\
\vdots & \vdots & \vdots & \vdots & \vdots & \vdots & \vdots & \vdots\\
h_7& -h_8 &h_9 & -h_{10}&h_{11}&-h_{12}&0&0\\
h_8&h_7&h_{10}&h_9&h_{12}&h_{11}&0&0\\
\vdots & \vdots & \vdots & \vdots & \vdots & \vdots & \vdots & \vdots\\
0 & 0 & h_{11} & h_{12} & -h_9 & -h_{10} & -h_7 & -h_8\\
0 & 0 & h_{12} & -h_{11} & -h_{10} & h_9 & -h_8 & h_7 \\
\end{array}
\right ]
\label{eq-30}
\end{equation}
where $h_i$, $i=1,2,\ldots,11$ and $h_j$, $j=2,4,\ldots,12$ are the
real and imaginary parts, respectively, of $h_{1,1}$, $h_{2,1}$,
$h_{3,1}$, $h_{1,2}$, $h_{2,2}$, $h_{3,2}$. The matrix $\check H^T$ is
$8\times 32$ where each row has 8 zeros, while each of the remaining
24 symbols has one of $h_1,h_2,\ldots,h_{12}$, repeated twice. 
Let's first ignore the repetition of $h_i$ in a row. Then,
the calculation of $\check H^T\check y$ takes $8\cdot
24=192$ real multiplications.  
The calculation of $\sigma=\check
h_1^T\check h_1=2\sum_{k=1}^{12}h_i^2$ takes $12+1=13$ real multiplications, In
addition, one needs 
4 real multiplications to calculate $\sigma^{-1}$, and 8 real 
multiplications to 
calculate $\sigma^{-1}\bar{\bar y}$. 
To calculate $\check H^T\check y$, one needs $8\cdot 23 = 184$ real
additions, and to calculate $\sigma$, one needs 11 real additions.
As a result, with this approach, one needs a total of 217 real multiplications
and 195 real additions to decode. 

For this example,
(\ref{eq-17}) specifies 300 real multiplications and 279 real additions. The
reduction is due to the elements with zero values in $\check H$.

It is important to make the observation that the repeated values of
$h_i$ in the columns of $\check H$, or equivalently $h_{m,n}^*$ in the
rows of $H^HA_k^H$ or $H^HB_k^H$, have a substantial impact on
complexity. Due to the repetition of $h_i$, by grouping the two values of
$\check y_j$ that it multiplies, it takes $8\cdot 12 = 96$ real
multiplications to compute $\check H^T\check y$, not $8 \cdot 24
=192$. The summations for each row of $\check H^T\check y$ will now be
carried out in two steps, first 12 pairs of additions per each $h_i$, and
then after multiplication by $h_i$, addition of 12 real numbers. This
takes $12 + 11 = 23$ real additions, with no change from the way the
calculation was made without grouping. With this change, the
complexity of decoding becomes 121 real multiplications and 195 real
additions, a huge reduction from 300 real multiplications and 279 real
additions. 

{\em Example 3:\/} We will now consider the code ${\cal G}_4$ from
\cite{tjc}. The parameters for this code are $N = K = 4$, $M=1$, and
$T=8$. It is given as
\begin{equation}
{\cal G}_4=\left[
\begin{array}{cccccccc}
s_1&-s_2&-s_3&-s_4&s_1^*&-s_2^*&-s_3^*&-s_4^*\\
s_2&s_1&s_4&-s_3&s_2^*&s_1^*&s_4^*&-s_3^*\\
s_3&-s_4&s_1&s_2&s_3^*&-s_4^*&s_1^*&s_2^*\\
s_4&s_3&-s_2&s_1&s_4^*&s_3^*&-s_2^*&s_1^*\\
\end{array}
\right]^T .
\label{eq-34}
\end{equation}
Similarly to ${\cal G}_3$ of Example 2, this code has the property
that ${\cal G}_4^H{\cal G}_4=2(\sum_{k=1}^K|s_k|^2)I_4$. 
The $\check H$ matrix is $16\times 8$ and can be
calculated as 
\begin{equation}
\check H = \left [
\begin{array}{cccccccc}
h_1 & -h_2 & h_3 & -h_4 & h_5 & -h_6 & h_7 & h_8 \\
h_2 & h_1 & h_4 & h_3 & h_6 & h_5 & h_8 & h_7 \\
h_3& -h_4 &-h_1 & h_{2}&h_{7}&-h_{8}&-h_5&h_6\\
h_4&h_3&-h_{2}&-h_1&h_{8}&h_{7}&-h_6&-h_5\\
\vdots & \vdots & \vdots & \vdots & \vdots & \vdots & \vdots & \vdots\\
h_5 & h_6 & -h_{7} & h_{8} & -h_1 & -h_{2} & h_3 & h_4\\
h_6 & -h_5 & -h_{8} & h_{7} & -h_{2} & h_1 & h_4 & -h_3 \\
\end{array}
\right ] .
\label{eq-35}
\end{equation}
This matrix consists entirely of nonzero entries. Each entry in a
column equals $\pm h_i$ for some $i\in \{1, 2,\ldots, 8\}$, every
$h_i$ appearing twice in a column. Ignoring this repetition for now,
calculation of $\check H^T\check y$ takes $8\cdot 16 = 128$ real
multiplications. Calculation of $\sigma$ takes 9 real multiplications, its
inverse 4 real multiplications, and the calculation of
$\sigma^{-1}\bar{\bar y}$ takes 8 real multiplications. Calculation
of $\check H^T\check y$ takes $8\cdot 15=120$ real additions, and 
calculation of $\sigma$ takes 7 real additions. 
As a result, with this approach, to decode, one needs
149 real multiplications and 127 real additions.

For this example, equation (\ref{eq-17}) specifies 156 real
multiplications and 135 real additions. The reduction is due to the
fact that one row of $\check H^T$ has each $h_i$ appearing twice. This
reduces the number of multiplications and summations to calculate
$\sigma$ by about a factor of 2.

However, because each $h_i$ appears twice in every row of $\check H^T$, the
number of multiplications can actually be reduced substantially.
As discussed in Example 2, we can reduce the
number of multiplications to calculate $\check H^T\check y$ by
grouping the two multipliers of each $h_i$ by summing them prior to
multiplication by $h_i$, $i=1,2,\ldots,8$. As seen in Example 2, 
this does not alter the number of real additions. With this simple
change, the number of real multiplications to decode becomes 85 and
the number of real additions to decode remains at 127.

{\em Example 4:\/} It is instructive to consider the code ${\cal H}_3$
given in \cite{tjc} with $N=3,$ $K=3$, $T=4$ which we will consider
for $M=1$ where
\begin{equation}
{\cal H}_3=\left[
\begin{array}{ccc}
s_1&s_2&{s_3}/{\sqrt{2}}\\
-s_2^*&s_1^*&{s_3}/{\sqrt{2}}\\
{s_3^*}/{\sqrt{2}}&{s_3^*}/{\sqrt{2}}&(-s_1-s_1^*+s_2-s_2^*)/{2}\\
{s_3^*}/{\sqrt{2}}&{-s_3^*}/{\sqrt{2}}&(s_2+s_2^*+s_1-s_1^*)/{2}\\
\end{array}
\right].
\label{eq-37}
\end{equation}
For this code, ${\cal H}_3^H{\cal H}_3=(\sum_{k=1}^3|s_k|^2)I_3$ is satisfied.
In this case, the matrix $\check H$ can be calculated as
\begin{equation}
\check H=\left[
\begin{array}{cccccc}
h_1&-h_2&h_3&-h_4&h_5/\sqrt{2}&-h_6/\sqrt{2}\\
h_2&h_1&h_4&h_3&h_6/\sqrt{2}&h_5/\sqrt{2}\\
h_3&h_4&-h_1&-h_2&h_5/\sqrt{2}&-h_6/\sqrt{2}\\
h_4&-h_3&-h_2&h_1&h_6/\sqrt{2}&h_5/\sqrt{2}\\
-h_5&0&0&-h_6&(h_1+h_3)/\sqrt{2}&(h_2+h_4)/\sqrt{2}\\
-h_6&0&0&h_5&(h_2+h_4)/\sqrt{2}&-(h_1+h_3)/\sqrt{2}\\
0&h_6&h_5&0&(h_1-h_3)/\sqrt{2}&(h_2-h_4)/\sqrt{2}\\
0&-h_5&h_6&0&(h_2-h_4)/\sqrt{2}&(-h_1+h_3)/\sqrt{2}\\
\end{array}
\right].
\label{eq-38}
\end{equation}
It can be verified that every column $\check h_i$ of $\check H$ has
the property that $\check h_i^T\check
h_i=\sigma=\|H\|^2=\sum_{k=1}^6h_k^2$ for $i=1,2,\ldots,6$. In this
case, the number of real multiplications to calculate $\check H^T\check y$
requires more caution than the previous examples. For the first
four rows of $\check H^T$, this number is 6 real multiplications per
row. For the last two rows, due to combining, e.g., $h_1$ and $h_3$ in 
$(h_1+h_3)/\sqrt{2}$ in the fifth element of $\check h_5$, and the
commonality of $h_5$ and $h_6$ for the first and third, and second and
fourth, respectively, elements of $\check h_5$, and one single multiplier
$1/\sqrt{2}$ for the whole column, the number of real multiplications
needed is 7. As a result, calculation of $\check H^T\check y$ takes 38
real multiplications. Calculation of $\sigma$ takes 6 real 
multiplications. One needs 4 real multiplications to calculate $\sigma^{-1}$,
and 6 real multiplications to calculate $\sigma^{-1}\bar{\bar y}$. First
four rows of $\check 
H^T\check y$ require 5 real additions each. Last two rows of $\check
H^T\check y$ require $4+7=11$ real additions each. This is a total of
42 real additions to calculate $\check H^T\check y$. Calculation of
$\sigma$ requires 5 real additions. Overall, with this approach one
needs 54 real multiplications and 47 real additions to decode. 

For this example, (\ref{eq-17}) specifies 66 real multiplications and 49
real additions. The reduction is due to the presence of the zero
entries in $\check H$. On the other hand, the presence of the factor
$1/\sqrt{2}$ in the last two rows of $\check H^T$ adds two real
multiplications to the total number of real multiplications.
\vspace{-3mm}
\section{Conclusion}
Equation (\ref{eq-17-}) yields the computational complexity of decoding
an OSTBC when its $\check H$ matrix consists only of nonzero entries in the
form of $h_i$ when $c=1$. It should be updated as specified in the paragraph
following (\ref{eq-17}) when $c>1$. 
The presence of zero values 
within $\check H$ reduces the computational 
complexity. In the examples its effect has been a reduction in the
number of real multiplications to calculate $\check H^T\check y$ by a
factor equal to the ratio of the rows of $A_k$ and $B_k$ that 
consist only of zero values to the total number of all rows in $A_k$ and
$B_k$ for $k=1,2\ldots,K$, with a similar reduction in the
number of real additions to calculate $\check H^T\check y$. 
With the modifications outlined above, (\ref{eq-17-}) specifies the
computational complexity of decoding the majority of OSTBCs.
In some cases, the contents of
the $\check H$ matrix can have linear combinations of $h_i$ values,
which result in minor changes in computational complexity as specified
by this formulation, as shown in Example 4. Finally, note that 
$L=2$ is a special
case where the signal belongs to one of the four quadrants,
calculation of and division by $c\|H\|^2$ are not needed and the
computational complexity will be correspondingly lower. 

\vspace{-4mm}
\bibliographystyle{IEEEtran}
\bibliography{IEEEabrv,bib/Ayanoglu} 

\end{document}